\documentclass[twocolumn,amscd, amsmath,amssymb,verbatim]{revtex4}

\usepackage{lgrind}        
\usepackage{chapterbib}    
\usepackage{color}         
\usepackage{graphics}      
\usepackage{graphicx}    
\usepackage{longtable}     
\usepackage{epsf}          
\usepackage{bm}            
\usepackage{thumbpdf}
\usepackage[colorlinks=true]{hyperref}  
\def\M{M}                                        
\def\mp{m_{p}}
\def\V{{ {\cal V}}}
\def\hV{{{\cal V}}}
\def\hVG{\hV_{{\cal N}_G}}
\def\hm{{m}}
\def\hn{{n}}
\newcommand{\be}{\begin{equation}}
\newcommand{\ee}{\end{equation}}

\begin{document}
\title{Can we live in the bulk without a brane?}
\author{Kurt Hinterbichler$^{1}$, Janna Levin$^{2,3}$, and Claire Zukowski$^{2,4}$}

\affiliation{${}^{1}$Center for Particle Cosmology, University of Pennsylvania,
Philadelphia, PA 19104}

\affiliation{${}^{2}$Institute for Strings, Cosmology and Astroparticle Physics, Physics Department
Columbia University, New York, NY 10027}

\affiliation{${}^{3}$Department of Physics and Astronomy, Barnard College,
New York, NY 10027}

\affiliation{${}^{4}$Department of Physics, University of California, Berkeley, CA 94720}


\begin{abstract}
We suggest a braneless scenario that still hides large-volume extra
dimensions. 
Ordinarily the strength of bulk gauge interactions would be diluted over the
large internal volume, making all the four dimensional forces weak. We use the fact that if the
gauge fields result from the dimensional reduction of pure
higher-dimensional gravity, then the strengths of the four dimensional gauge
interactions are related to the sizes of  corresponding cycles averaged over
the compact internal manifold. Therefore, if 
a gauge force is concentrated over a small cycle it will not be
diluted over the entire manifold. Gravity, however, remains diluted over
the large volume. Thus large-volume, large mass-gap extra
dimensions with small cycles can remain hidden and result in a
hierarchy between gravity and the other forces. However,
problematically, the cycles are required to be smaller than the
higher-dimensional Planck length and this raises concern over quantum gravity corrections. We speculate on
possible cures.
\end{abstract}

\maketitle

A low-budget yet concrete observation about our universe is that it
is four-dimensional, and although human beings by the billions confirm this observation
daily, it might not be true. Our universe may have multiple extra
spatial dimensions and only {\it appear} to be
four-dimensional. 
Extra dimensions could hide if they are curled up so
small that no observations to date could excite modes energetic enough
to probe these
directions. Originally, interest in large-volume extra dimensions inspired
braneworlds as a new means to hide the extra dimensions -- float our
universe on a 3-brane and confine all Standard Model fields to that
braneworld
\cite{ArkaniHamed:1998rs,Antoniadis:1998ig,Randall:1999ee}.
In this article, we describe a braneless
alternative that allows us to hide large-volume extra dimensions. The internal
manifolds have in common three essential features: (1) large volume,
(2) lowest modes that are energetically
expensive despite the large volume, and as we will see (3) some small
Killing cycles.  However,
the small cycles are {\it very} small and therefore
penetrate quantum gravity scales as we discuss shortly.

We can lend intuition for why these three features are essential. Part of
the picture was presented in a
previous article \cite{Greene:2010ch}. It is commonly assumed that the
larger the volume, the 
lower the natural harmonics on the space.
The lower the notes, the less the effort that is required to play them. 
If we were not
confined to a brane,
we
should expect to have observed a large internal volume already.
However, this expectation is contradicted
by an infinite number of known manifolds that despite their large
volume have no low notes
\cite{Kaloper:2000jb,Starkman:2000dy,Starkman:2001xu,Nasri:2002rx,Greene:2010ch}. Such
spaces would remain hidden despite their large size because it remains
too energetically expensive to probe them. 

Still, an expensive spectrum of modes is not sufficient to
free the world from incarceration on a brane. 
If Standard Model fields were allowed to live in the
bulk, it might seem that the strength of all the gauge interactions would be diluted
over the large volume -- just 
as the strength of gravity would be diluted -- and all forces would be weakened relative to the true
fundamental scale. For this reason, Standard Model gauge fields were
localized on a brane while the
gravitational field inhabited the bulk.
This split between habitats enforced a hierarchy between gravity
and particle physics \cite{ArkaniHamed:1998rs,Antoniadis:1998ig,Randall:1999ee}.

If we are to do away with the brane altogether and allow all fields to
fill the bulk, then we need to save the gauge couplings
from dilution over the large internal volume. We show here that this is
possible if the gauge fields are generated by the dimensional
reduction of a purely gravitational field -- as in the original
Kaluza-Klein reduction \cite{Kaluza,Klein} -- and there are some small
Killing cycles around the internal volume. In this picture, a
photon is really a metric oscillation around an $S^1$, and the strength of
its coupling is inversely proportional to the size of that cycle. So
while the volume is large, electromagnetic interactions involve
only one small circle and not the entire manifold. 
The dilution over the internal volume is
compensated by localizing all gauge interactions over small cycles,
instead of confining them to a brane.

We will review the dimensional
reduction of gravitational fields from a higher-dimensional universe
down to a $4$-d universe and show that
the conditions we want our manifold to satisfy are the following: 

\textbf{\textit{Large Volume}}: Under dimensional reduction, the 
integrated volume ${\cal V}$ (in units of $M^N$) 
is related to the ratio of the
observed $4$-d Planck mass, $\mp$, to the
fundamental higher-dimensional Planck scale, $M\sim {\rm TeV}$, through
\be
{\cal V}=\left (\frac{\mp}{ \M}\right )^2 \quad .
\label{Eq:one}
\ee
This only generates a hierarchy if the Higgs mass is small compared to $\mp$.
Since the $4$-d metric is not warped, any bulk scalar field
added by hand will automatically have the mass it did in the bulk
\cite{ArkaniHamed:1998rs,Antoniadis:1998ig}. Although vacuum expectation values and couplings will be
affected, combinations of them lead to invariant masses
\cite{Greene:2010ch}. So if the Higgs is a bulk scalar field of mass
$M$ in the bulk, it will reduce to a 4-d scalar field of mass $M$.

\textbf{\textit{Large Mass Gap}}:
The mass gap, set by the minimum non-zero eigenvalue of the Laplacian on the
internal manifold, must be large to suppress Kaluza-Klein excitations,
a condition we express as,
\be
m_{KK} \gtrsim M.
\label{Eq:two}
\ee

\textbf{\textit{Small Cycles}}:
Some of the $4$-d gauge couplings, i.e. those of the Standard Model, must be of order one. The $4$-d gauge couplings can
be expressed as 
\be g_{4d} \sim \left({\langle s^2\rangle^{1/2} M\V^{1/2}}\right)^{-1},\ee
where $\langle s^2 \rangle^{1/2} $ is the root mean square of the circumference of the corresponding Killing cycle
over the internal space (as we
review below \cite{Weinberg1983265}).
Setting this $\simeq 1$ gives
\be
\langle  s^2 \rangle ^{1/2} \sim (M\V^{1/2})^{-1} \quad .
\label{Eq:three}
\ee
When all three conditions are met, we have
large-volume extra dimensions that can be hidden without invoking a
brane while still affecting a hierachy between the weakness of gravity
relative to particle interactions.

However, problematically, by Eq.\ (\ref{Eq:one}) in Eq.\ (\ref{Eq:three}),
\be
\langle  s^2 \rangle ^{1/2} \sim \mp^{-1} \quad .
\label{Eq:threebad}
\ee
The cycles corresponding to gauge interactions are smaller by a factor of the hierarchy than the
higher-dimensional Planck length, leading to curvature invariants (or
analogous topological invariants) that are large and
susceptible to uncontrolled quantum corrections.
We will consider internal manifolds that are a direct product of
submanifolds as well as internal manifolds that are warped products of
submanifolds. Although the internally warped spaces seem promising in
that the small cycles are of order $M^{-1}$, the warping shrinks the
cycles over the span of the manifold so they are metrically small in
places. Consequently, we run into trouble with large curvature
invariants and cannot claim controlled quantum gravity corrections.

\bigskip
\centerline {\textbf{Gravity Reduction}}

The gravity reduction of Kaluza and Klein
\cite{Kaluza,Klein} provided a remarkable, explicit demonstration of unification: a metric
flux around a circle in $5$-d appeared to the $4$-d world as the
photon. Since then, all manner of gauge groups have been shown to
result from the dimensional reduction of pure gravity over
higher-dimensional manifolds. 
We consider 
pure  ($4+N$)-d gravity
\be S = \frac{\M^{2+N}}{2}\int d^{4+N}x \sqrt{-G} R(G), \ee
on a product space
$\mathcal{M}\times\mathcal{N}$, where $\mathcal{M}$ is 
$4$-d and the internal $N$-d
manifold $\mathcal{N}$ has isometry group ${\cal G}$.
Let us consider only zero-modes under dimensional reduction, which is equivalent to
assuming that the Kaluza-Klein excitations of the metric can be
ignored at the energy scales we are considering. Then we use the
ansatz for the metric \cite{PhysRevD.12.1711, Weinberg1983265} 
\be
G_{AB}=\left( \begin{array}{c|c} g_{\mu \nu} + A_\mu^i A_\nu^j
  \xi_{i}^m \xi_{j}^n\hat{g}_{\hm\hn} &
  A_\mu^i\xi_{i\hn}\\ \hline
A_\nu^j \xi_{j\hm} & \hat{g}_{\hm \hn}\end{array}\right),
\label{Eq:GAB}
\ee
where $\mu=0,...,3$ runs over $4$-d coordinates $x$ and $\hm=5,...,4+N$
runs over the internal coordinates $y$.
The $\xi_i^m(y)$ are the Killing vectors of ${\cal N}$ that under the Lie bracket obey the algebra of the isometry group of the internal manifold, 
\be [\xi_i, \xi_j] ^\mu=  C_{ij}^{\ \ k} \xi_k^\mu \quad ,\ee
with $C_{ij}^{\ \ k}$ the canonical ($\sim1$) structure constants of the algebra.
In words, the internal spacetime symmetries appear to us in $4$-d to be proper
gauge symmetries, and the off-diagonal excitations of the metric
camouflage as gauge fields.

After dimensional reduction, 
we get $4$-d gravity with metric $g_{\mu\nu}$, $4$-d Yang-Mills gauge fields $A_\mu^i$ with gauge
group isomorphic to ${\cal G}$, and scalar field moduli from the
internal metric $\hat g_{\hm \hn}$,
\begin{align} 
\int d^{4+N} x\sqrt{-G} \frac{M^{2+N}}{2}\left[ \right.  R(g)
 \left. - \hat g_{\hm\hn}\xi_i^{\hm}\xi_j^{\hn}
  \frac{1}{4}F^i_{\mu \nu}F^{j\mu \nu}\right]
+ ...
\end{align}
 where $F^i_{\mu \nu}=\partial_\mu A_\nu^i-\partial_\nu A_\mu^i+ C_{jk}^{\ \ i}A_\mu^jA_{\nu}^k$  is the standard non-abelian Yang-Mills curvature.  The $...$ indicates additional moduli terms, including
 curvature terms like
$R(\hat g)$ that serve as a
potential for the moduli of the internal dimension and/or contribute
to the cosmological constant. (While both moduli
stabilization and the cosmological constant are important problems for any higher-dimensional model,
we defer to the rich literature on the subjects.)

Integrating the Einstein-Hilbert term over $y$, we see that the $(4+N)$-d Planck constant is related to to $4$-d Planck constant by 
\be
{\cal V} \equiv M^N\int d^N y\sqrt{\hat{g}}  =\left (\frac{ \mp}{\M}\right )^2 \ . 
\label{Eq:one_2}
\ee
This is the first condition,
Eq.\ (\ref{Eq:one}). The internal volume under dimensional reduction
must be large relative to the fundamental scale $M$.

For a given simple part of the gauge group, the kinetic coefficient of the gauge fields can be chosen diagonal, and its coefficient will be related to the 4-d gauge coupling $g_{4d}$, 
\be
\frac{M^{2}\V}{2}\langle\hat{g}_{\hm \hn}
\xi_i^{\hm} \xi_j^{\hn}\rangle
={1\over g^2_{4d}}\delta_{ij} \ ,
\label{Eq:normKV}
\ee
where $\left < \varphi(y) \right >=M^N\V^{-1}\int d^N y \sqrt{\hat{g}} \varphi(y)$
indicates an average of a function $\varphi(y)$
over the internal volume.

The action is then the  canonical action for 4-d
gravity with 4-d gauge fields:
\begin{align} 
S = \int d^{4} x\sqrt{-g} \left[ \frac{\mp^2}{2} R(g) -
  \frac{1}{4g_{4d}^2}\left( F^i_{\mu \nu}\right)^2\right]
+ ...
\end{align}

The massive gravitons
corresponding to
Kaluza-Klein modes for the metric that would
appear in the action have masses $\sim m^2_k$ corresponding to
eigenvalues of the scalar Laplacian on the higher-dimensional manifold,
\be
\nabla_{(N)}^2 \psi_k= - m_k^2 \psi_k.
\ee
To suppress Kaluza-Klein modes, 
we require non-zero eigenvalues of the
Laplacian to be large,
\be
m_{k_{\rm min}}\gtrsim M \ .
\ee
This is the second condition, Eq.\ (\ref{Eq:two}).

To obtain the third condition, Eq.\ (\ref{Eq:three}), we
give a minimal review of the argument in \cite{Weinberg1983265}, to show that 
gauge coupling constants in the lower dimensional theory, given by Eq.\ (\ref{Eq:normKV}), can be interpreted as averaged circumferences over the compact internal
manifold. Given a compact, simple Lie group acting on a compact manifold, there is a Killing vector $\xi_i^m$ corresponding to each Lie algebra generator $T_i$.  
A generic Killing vector $\xi^m$ generates closed orbits $Y^m(\lambda)$, parametrized by some $\lambda$, in the corresponding compact manifold,
\be
{d\over d \lambda} Y^m(\lambda)=\xi^m(Y(\lambda)).
\label{Eq:orbit}
\ee
Given a starting point $y$ for the orbit, the solution is
\be Y^m(\lambda,y)=e^{\lambda\xi^n(y)\partial_n}Y^m(0,y),\ee
where the partial derivative in the exponent is with respect to $y$.  This generates the exponential map on the group manifold, and since the generators are normalized canonically (structure constants are $\sim 1$), and the group is compact, the curve comes back to its starting point after some order one range of $\lambda$ (usually $\lambda_{\rm max}=2\pi$).

The circumference of this curve is thus
\be s(y)=\int_0^{\lambda_{\rm max}}d\lambda \sqrt{\hat g_{mn}(Y){dY^m \over d \lambda} {dY^n \over d \lambda}}.\ee
By differentiating the quantity in the square root with respect to $\lambda$, using Eq.\ (\ref{Eq:orbit}), and using Killing's equation ${\cal L}_\xi \hat g_{\mu\nu}=0$, it is straightforward to check that the integrand is actually independent of $\lambda$, so we have 
\be s(y) \sim \sqrt{\hat g_{mn}\xi^m\xi^n}.\ee

Taking the average as $y$ varies over the submanifold ${\cal N}$, we have from Eq.\ (\ref{Eq:normKV})
\be 
g \sim \frac{1}{M\V^{1/2} \sqrt{\left < s^2\right >}} \ .
\label{Eq:it}
\ee

The argument also generalizes to the case of a $U(1)$ group factor, though one has to couple in matter to read off the coupling strength \cite{Weinberg1983265}.  Weak or strong gauge couplings correspond to
large or small cycles respectively. 

\bigskip
\centerline {\textbf{Direct Product Spaces: No Warping}}

Consider as a first example a
direct product of $4$-d ${\cal M}$ with
an $N$-dimensional manifold built from a string
$S^n\times...\times S^n$, of $D$ hyperspheres of radius
$R_1$ \cite{Greene:2010ch} taken in product with one much smaller $n$-sphere of radius $R_2$, so $N=n(D+1)$.  The gauge fields will come from the smaller sphere, and we have $\langle s^2\rangle ^{1/2}\sim R_2$.
By our 3 conditions, the internal space is subject to the constraints:
\begin{align}
& {\cal V} ={\cal V}_{S_1}^D {\cal V}_{S_2}\sim (R_1M)^{Dn}(R_2M)^n\sim (\mp/M)^2, \nonumber \\
& m_{KK} \sim R_1^{-1} \sim M, \nonumber \\
& g_2 \sim (R_2 M{\cal V}^{1/2})^{-1} \sim (R_2 \mp)^{-1}\sim 1  \quad.
\end{align}
Choosing $R_2\sim \mp^{-1}$, $R_1 \gtrsim M^{-1}$ and $D \gg 1$ easily
satisfies all 3 conditions.
The gauge fields from $R_2$ couple with strength $g_2\sim 1$ while
the string of large $S^n$'s will couple with a strength suppressed by a factor
of $\M/\mp \sim 10^{-16}$.

In general we can build the internal manifold as a product of any
large-volume, large mass gap space with highly diluted gauge couplings
times a small manifold with undiluted gauge couplings. 
Another interesting internal manifold is provided by a squashed
$T^2$ in product with a small space. 
Unlike the previous example, this manifold does not require
large dimensionality. The large volume, large mass gap comes from the
squashed $T^2$ as was shown in
\cite{Dienes:2001wu,Dienes:2001ac} while the undiluted gauge
coupling could come from a small 
internal manifold
such as $\mathbb{CP}^2\times\mathbb{S}^2\times\mathbb{S}^1$, which has
the isometries of the Standard Model gauge group 
\cite{Witten1981412}.

There are an infinite number of 2-surfaces that could participate in
this construction. In \cite{Greene:2010ch}, 
compact hyperbolic 2-surfaces were considered.
Surfaces of arbitrarily large genus, and
therefore arbitrarily large area,
$A=4\pi (g-1)R^2$,
were shown to have
minimum
eigenvalue
of roughly $k_{\rm min} \sim 1/(2R)$ \cite{Buser1,Buser2,Selberg,LRS,BrooksMakover1,BrooksMakover2}.
With $R\sim M^{-1}$ and $g\gg 1$, these qualify as large mass-gap, large-volume
manifolds. 
Additionally, hyperbolic spaces have no Killing vectors and so would
not contribute any additional, unwanted gauge fields.
We could equally well take the internal space to be a product of these
large-volume, large mass-gap manifolds with a small
internal manifold
whose
isometries generate the Standard Model.

These direct product spaces have a nice interpretation: There is a
large internal volume but gauge
fields correspond to excitations along small cycles and so do not require
ringing the whole big manifold, just a small piece of it. Therefore
the gauge coupling is not diluted, while gravity is.

Despite this nice interpretation, these examples are flawed.
One
of the geometric scales, $\mp^{-1}$, is many orders of magnitude
smaller than the fundamental length scale, $\M^{-1}$, with the considerable
disadvantage that small cycles might force us into quantum gravity
arenas, undermining the consistency of the analysis.  The concern is
that higher-dimensional operators of the form
$R^2,R_{\mu \nu}R^{\mu \nu}...$ become significant.  Before we speculate on possible resolutions,
we turn to warped {\it internal} spaces next, although we will see that these
also have the problem of 
small cycles.

\bigskip
\centerline {\textbf{Warped Internal Spaces}}

Consider now a product of $\mathcal{M}$ with an internal space
$\mathcal{N}= H^2\times {\cal N}_G$ that is a product of a swath of the
hyperbolic plane, $H^2$, with coordinates $y,z,$
and a space ${\cal N}_G$ with coordinates $\hat x^m$ and metric $\hat g_{\hm \hn}$ with isometry group the desired gauge
group. Let ${\cal N}_G$ carry a 
warp factor $f(y) $ dependent on only the $y$ coordinate of the hyperbolic
plane (note that the warping is internal and that the $4$-d metric carries no warp factor),
\be
ds^2=g_{\mu \nu}dx^\mu dx^\nu+f(y)\hat g_{\hm \hn}d\hat x^{\hm}d\hat
x^{\hn} +{L^2\over y^2}\left(dy^2+dz^2\right) \quad .
\ee
We have used the upper half-plane representation of the hyperbolic
plane, with $L$ the curvature scale. We take a square swath on the plane between the limits $0< z < 1$
and $\epsilon < y< 1$ where $\epsilon$ is small but non-zero in
order to render lengths and areas finite.

In order for all components of the higher-dimensional metric to
transform properly under the gauge transformations and create the
illusion of gauge bosons, we require the
Killing vectors that generate them to be Killing vectors of the 
{\em entire} metric, not just of the submanifold ${\cal N}_G$.  But because the warp factor depends only on the coordinates of $H^2$, Killing vectors of ${\cal N}_G$ are automatically Killing vectors of the entire internal manifold. 
Finite volume hyperbolic manifolds do not have Killing vectors and so 
do not introduce additional gauge fields. 

Choosing $f(y)=\alpha y $, with $\alpha$ some order one constant, and 
a submanifold ${\cal N}_G$ of dimension 2 and un-warped volume
$\hVG\sim1$ in units of $M^2$, 
the $4$-d Planck mass by Eq.\ (\ref{Eq:one_2}) is
\begin{align}
\frac{\mp^2}{M^2}& =\hVG M^2L^2\int_0^1dz\int_\epsilon^1dy \ y^{-2} f(y) \nonumber \\ & = \alpha
\hVG(LM)^{2} \ln(1/\epsilon).
\end{align}
The
$4$-d Planck mass
is then large if we take $\epsilon$ very near zero.
We thereby meet condition Eq.\ (\ref{Eq:one}) to create a weakened
strength of
gravity compared to a fundamental scale.

To check the mass gap of Eq.\ (\ref{Eq:three}), we note that the KK
modes on ${\cal N}_G$ should be of order $M^{-1}$. Also, the
eigenmodes on
the swath of the hyperbolic plane should be subcurvature modes and
therefore the eigenvalues are bounded from below by $1/(2L)$ so 
the mass gap is comfortably large for $L \sim M^{-1}$.

The gauge couplings are set by the normalization of the Killing vectors, 
Eq.\ (\ref{Eq:normKV}). Due to the factor of $\hat g_{\hm\hn}$ in that inner product, 
an additional factor of $f$ is introduced so that the
gauge couplings consistent with Eq.\ (\ref{Eq:it})
are set by $\V{\left <s^2\right >}=\hVG R^2M^2L^2\int (f^2/y^2) dzdy=\alpha^2\hVG(LM)^2 R^2(1-\epsilon)$ where $R$ is the
characteristic size of the un-warped cycle on the submanifold ${\cal N}_G$. It follows that
\be 
{g} \sim \left (\alpha M^2 LR\hVG^{1/2} \right )^{-1} \quad .
\ee
For $\hVG\sim LM\sim RM \sim \alpha\sim 1$, we have an $\mathcal{O}(1)$
coupling.

The hierarchy between fundamental scales has been shifted to a
hierarchy between geometric scales. 
As with the direct product, the warped internal product dilutes
gravity over a large internal volume while gauge fields correspond to
excitations over small cyles. 
Unfortunately, also like the direct
product, the cycles vary by $\sqrt{f(y)}R$ and are metrically small in places.
The hazards of quantum gravity thus reappear as the
bulk Ricci scalar on the internal space is larger by the warp factor
than is tolerable, with a contribution of the form $f^{-1}R(\hat{g})$.

\bigskip
\centerline {\textbf{Speculation}}

It may be that there is a no-go theorem that ensures the small cycles
we need are always catastrophically small if there are no branes. 
On the other hand, although we did not present the calculation here, we find the dimensional reduction
of pure gravity where the external $4$-d space {\it is} warped as in
Randall-Sundrum -- as opposed to the internal warping detailed above -- does
allow for gauge fields to live in the bulk with order unity gauge
couplings following this prescription. However, this construction is less novel, and furthermore, the hierarchy requires the
Higgs to be constrained to a brane so does not qualify as braneless.

Finally, it is a celebrated result of string theory that small cycles are controlled in the
UV theory as extra light degrees of freedom resurface and
naturally resolve any metrical divergences \cite{Strominger:1995cz}. We
speculate that in a stringy formulation, a similar resolution
may allow a fully braneless model with small cycles and good gauge
couplings.  If string theory is the UV completion, there is also the worry that winding modes around a homotopically non-trivial small cycle can become light, since their mass $\sim R/ \alpha'$.  This is not a problem if the Killing cycles are homotopically trivial, as in our example with spheres $S^n$.
We also mention that there are 
explicit examples of
hyperbolic 3-folds with large volume and small geodesics 
(see Snappea \cite{snappea}). The curvature
invariants could be stabilized at $M^{-1}$ while sustaining large
volume and small cycles. Additional operators based on
curvature invariants
would be controlled, skirting the problem of small cycles.

There are of course other issues that must traditionally be confronted in any Kaluza-Klein scenario, such as the incorporation of chiral fermions and stabilization of moduli.  In the meantime, 
it is encouraging that a braneless cosmos might hide large-volume extra dimensions.
While all fields are smeared out over the large-volume, interactions
with gauge fields
are concentrated over relatively small cycles and thereby
manage to remain undiluted. 
It is intriguing to imagine that we live smeared out
over a large drum and our experience of forces other than gravity are
an illusion created by the cadence of
small hidden subspaces.

\bigskip
{\bf Acknowledgements}: 
We thank  Gregory Gabadadze, Brian Greene, Daniel Kabat, David Kagan, Alberto Nicolis, Massimo Porrati, and
Dylan Thurston for invaluable conversations. 
KH acknowledges support from NASA ATP grant NNX08AH27G, NSF grant PHY-0930521, DOE grant DE-FG05-95ER40893-A020 and the University of Pennsylvania. JL acknowledges support from an NSF Theoretical Physics grant,
PHY - 0758022. JL also gratefully acknowledges
a KITP Scholarship. 

\bibliography{InternalWarped}

\end{document}